\documentclass[letterpaper, 10 pt, conference]{ieeeconf}  

\IEEEoverridecommandlockouts                              

\usepackage{cite}

\usepackage{graphicx} 

\title{\LARGE \bf
Multimodal Speech Emotion Recognition Using Modality-specific Self-Supervised Frameworks
}

\author{Rutherford Agbeshi Patamia$^{1}$, Paulo E. Santos$^{2}$, Kingsley Nketia Acheampong$^{1}$,\\ Favour Ekong$^{1}$, Kwabena Sarpong$^{1}$ and She Kun$^{1}$$^{*}$
\thanks{$^{*}$Corresponding Author: \it\small kun@uestc.edu.cn}
\thanks{$^{1}$R. A. Patamia, K. N. Acheampong, F. Ekong, K. Sarpong and S. Kun are with School of Information and Software Engineering, University of Electronic Science and Technology of China, Chengdu, SC 610054, China.
        {(Emails: \it\small{rap@std.uestc.edu.cn, kingsleyna@uestc.edu.cn, favourekong@std.uestc.edu.cn, t.shele@outlook.com and kun@uestc.edu.cn})}}%
\thanks{$^{2}$Paulo E. Santos is with the Centre for Defence Engineering, Research and Training College of Science Engineering, Flinders University,
        Tonsley, SA 5042, Australia.
        {(Email: \it\small{paulo.santos@flinders.edu.au})}}%
}

\begin{document}
\textbf{IEEE Copyright Notice}\\

Copyright (c) 2023 IEEE\\

Personal use of this material is permitted. Permission from IEEE must be obtained for all other uses, in any current or future media, including
reprinting/republishing this material for advertising or promotional purposes, creating new collective works, for resale or redistribution to servers or lists, or reuse of any copyrighted component of this work in other works.\\

\textbf{Accepted to be published in:} IEEE SMC-2023; October 1-4, 2023 - Honolulu, Oahu, Hawaii, USA. https://ieeesmc2023.org/

\maketitle
\thispagestyle{empty}
\pagestyle{empty}

\begin{abstract}

Emotion recognition is a topic of significant interest in assistive robotics due to the need to equip robots with the ability to comprehend human behavior, facilitating their effective interaction in our society. Consequently, efficient and dependable emotion recognition systems supporting optimal human-machine communication are required. Multi-modality (including speech, audio, text, images, and videos) is typically exploited in emotion recognition tasks. Much relevant research is based on merging multiple data modalities and training deep learning models utilizing low-level data representations. However, most existing emotion databases are not large (or complex) enough to allow machine learning approaches to learn detailed representations. This paper explores modality-specific pre-trained transformer frameworks for self-supervised learning of speech and text representations for data-efficient emotion recognition while achieving state-of-the-art performance in recognizing emotions. This model applies feature-level fusion using nonverbal cue data points from motion capture to provide multimodal speech emotion recognition. The model was trained using the publicly available IEMOCAP dataset, achieving an overall accuracy of 77.58\% for four emotions, outperforming state-of-the-art approaches
\end{abstract}

\section{INTRODUCTION}\label{sec1}

Emotions are a distinguishing feature of how humans interact with each other on a personal level. Emotions either introduce or eliminate ambiguity in communication, changing the meaning of what is being conveyed. Because emotions are essential in human-to-human communication, recent research has sought to replicate similar characteristics in machines, enabling them to be more effective in comprehending and communicating with people\cite{RB}. Human emotion recognition by machines aids various research fields, such as virtual reality, gaming, robotics, and customer care operations. For instance, if an automated call center system is able to infer a customer's emotional state, the application may be able to provide more appropriate responses or send the call to a human operator straightaway \cite{1}. Emotion detection systems can be used in virtual reality and games to detect a player's emotional suffering, paving the way for more realistic, engaging, and immersive gaming experiences \cite{CC2}. Recognizing human emotion by a robot can also lead to more natural human-robot interactions \cite{CC1}. 

Over the years, research into emotion recognition has been performed by analyzing speech from low-level acoustic information. A machine-learning classifier is usually applied to map this input information to a specific emotion category. With the recent advancement of deep learning methods \cite{2,3,4}, multiple efforts have been undertaken to learn emotion representations from audio signals using neural networks. However, the development of deep learning-based systems is frequently hampered by a lack of labeled data. In contrast to automated speech recognition (ASR) datasets, commonly used Speech Emotion Recognition datasets \cite{IE5,6,7} are restricted in complexity and small in size. Furthermore, systems trained on these datasets may not be generalizable to other domains, such as customer service centers\cite{1}. As a result, self-supervised pre-trained models such as wave-to-vector (wav2vec)\cite{21} and Bidirectional Encoder Representations from Transformers (BERT)\cite{28} have been created to address the above-mentioned issue by learning from large-scale audio and text datasets without the need of extensive labeling.

Transformer-based models are recognized to solve the difficulties mentioned above. Since most pre-existing transformer-incorporated models utilized in this field are trained on large volumes of unlabeled data, and are only fine-tuned on labeled data for emotion recognition tasks. They do not require large datasets, with detailed labeling, during training in contrast to other deep learning models. Studies have also demonstrated that the features derived by these novel transformer-based algorithms outperform classic spectral-based features on Speech Emotion Recognition (SER) \cite{11}. 

The following are the contributions of this work:
\begin{enumerate}
    \item The exploration of fine-tuning dual modality-specific transformer-based pre-trained architectures (using high-level speech audio and text representations as input) for multimodal emotion recognition;
	 \item The inclusion of nonverbal behavioral cues from the Motion Capture (MoCap) spectrum, which is universal and independent of cultural background, to create a versatile, robust, fair, and efficient multimodal emotion recognition system requiring low computational resources;
	\item The investigation of how a simple fusion method for speech, text, and MoCap modalities may outperform more complex approaches  when using Self Supervised Learning models with similar architectural properties.
\end{enumerate}
The proposed framework is a modular and extensible multimodal speech emotion recognition system. It consists of independently trainable sub-modules for speech, text, and motion capture emotion recognition. Each sub-module can be trained separately, and the absence of any modality or sub-module does not harm the overall framework. The pre-final layers must only be fine-tuned for combining the sub-modules, avoiding time-consuming complete model retraining. The framework leverages the advanced properties of pre-trained models to overcome accuracy loss and feature explosion during the feature extraction process.


\section{RELATED WORK}\label{sec2}
Speech Emotion Recognition (SER) is a technique for recognizing and classifying a speaker's emotion from audio inputs. Neural network-based solutions have positively impacted and continue to influence contemporary advancements in SER studies. Many researchers have adopted deep learning algorithms for robust feature representation in different domains, and SER is no exception. This development can be attributed to the fast pace of solutions to visual task recognition \cite{16}. In \cite{12}, the authors introduced a comprehensive model incorporating convolutional layers and a multi-head self-attention mechanism, which utilized deep encoded linguistic information and audio spectrogram representation to perform emotion recognition in speech. Also, to address the class imbalance in their dataset, they carried out down-sampling and ensembling, further improving the SER accuracy. \cite{13} presented a paradigm for essential sequence segment selection based on a Radial-Based Function Network (RBFN) with cluster similarities. These features were then sent into a deep Bidirectional Long Short-Term Memory (BLSTM), which learns the temporal information required to recognize the ultimate state of emotion. An RBF-based K-mean clustering algorithm was implemented in conjunction with the CNN model to extract more valuable features from spectrograms of the speech signal to achieve precise recognition performance. Consequently, a CNN \& LSTM model with an attached attention model was used to investigate SER theories in \cite{15}. Additionally, the significance of acoustic context information was investigated, which provided the means to verify that even relatively insignificant acoustic cues encompass emotional information valuable to the overall project.

Leonardo et al. \cite{22} developed a shallow neural network using transfer learning to extract features from pre-trained wav2vec 2.0 models for the SER problem. By merging different layers in the wav2vec 2.0 model with the trainable weights learned concurrently with the downstream model, they found that integrating information from multiple levels resulted in better outcomes than using only the encoder output, as observed in earlier studies. In a subsequent study, the authors \cite{23} presented two techniques, vanilla fine-tuning (V-FT) and task adaptive pretraining (TAPT), that utilize wav2vec 2.0 for SER. They also introduced a pseudo-label task adaptive pretraining (P-TAPT) approach that addresses the mismatch challenge between pretraining and the target domain by continuing the relevant process on the target dataset and learning contextualized emotion representations, resulting in improved overall model performance. 

The paper presents a new approach to multimodal speech emotion recognition that combines three modalities, including a nonverbal data point independent of cultural background, using feature concatenation. Unlike previous work, the modality components are treated independently, which means that the absence of one modality would not disrupt the entire system. In such a scenario, only the precursive layer of the system would need to be retrained, not the other modalities.


\section{METHODOLOGY}\label{sec3}
\begin{figure*}[ht!]
\begin{center}
     \includegraphics[width=15cm]{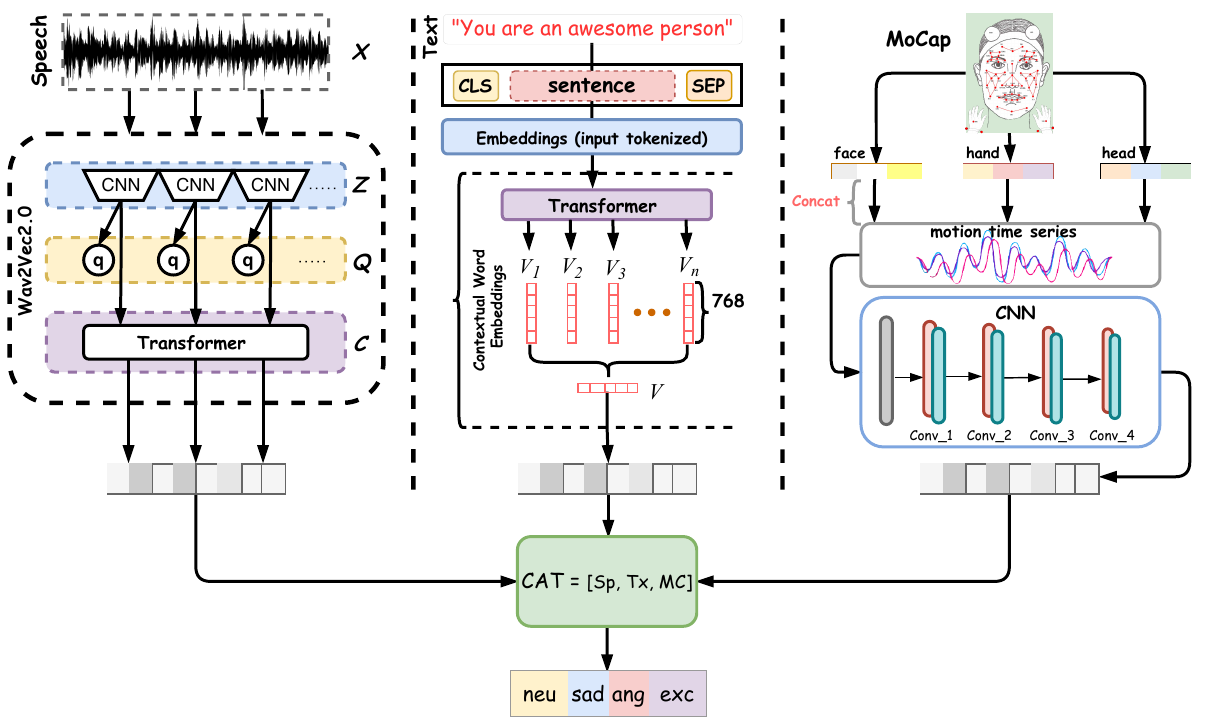}
\end{center}
\caption{The structure of our multi-modal model.}
\label{fig:final_model}
\end{figure*}

We explored the use of modality-specific self-supervised pre-trained models plus the addition of a  non-verbal data point (Motion Capture) for the task of multimodal speech emotion recognition. This consists of wav2vec2.0 (pre-trained on speech corpus), BERT (pre-trained on text corpus), and CNN mechanism for the non-verbal data point. To guarantee a competently informed multimodal emotion fusion, the outcomes of all modalities were merged at the feature level using user-defined criteria. Fig.~\ref{fig:final_model} illustrates the detailed implementation of our concept. It should be noted that the fused characteristics were subsequently routed via a fully connected layer for a decision or predicted emotional class.

\subsection{Wav2Vec2.0 for Speech Modality}\label{subsec2}
Wav2vec2.0 is a self-supervised pre-trained audio representation learning system. Its design involves a multi-level convolutional feature encoder f: \textit{X → Z} that accepts raw audio \textit{X} as input and generates latent speech embeddings $Z_{1},...,Z_{T}$ for \textit{T} time-steps. The encoder is made up of many blocks, the first of which is a temporal convolution succeeded by layer normalization\cite{26} and the last of which is a Gaussian error linear units (GELU) activation function\cite{27}. The encoder's total stride specifies the number of time-steps \textit{T} sent to the Transformer g: \textit{Z → C} to construct representations $C_{1},..., C_{T}$ aggregating data from the entire sequence\cite{28}. The output of the feature encoder is discretized into $qt$ by applying a quantization module \textit{Z → Q} to encode the targets, which involves distinguishing the genuine quantized latent speech representation for a masked time step from a set of distractors.

We adopted wav2vec2.0 as the acoustic encoder for speech emotion recognition by fine-tuning it with labeled emotion-dependent data. A linear projection layer was added on top of contextual representations to translate them into emotion-dependent classes. The model was improved by fine-tuning the CTC loss \cite{18}, and the identified emotion-dependent components were mapped to an emotion category using a majority vote for speech emotion recognition.

\subsection{BERT for Text Modality}\label{subsec3}
The text-related aspect of the study involved using deep contextual embeddings (BERT)\cite{28} to calculate the vector representation of words. BERT is a pre-trained model with a Transformer encoder architecture\cite{28} that uses a multi-layer Transformer encoder, with each layer consisting of a multi-head self-attention sublayer accompanied by a position-wise fully linked feedforward network. The model employs an embedding table and a final fully connected layer to convert hidden vectors to an output softmax over the vocabulary. To achieve the desired results, the study fine-tuned a pre-trained BERT model with 12 transformer blocks, 12 attention heads, and 110 million parameters.

\subsection{CNN-LSTM with Self Attention for Motion Capture Modality}\label{subsec4}
The use of nonverbal data, specifically motion capture (MoCap), has been recognized as crucial for emotion analysis in human-centered cognitive computing. MoCap technology is capable of tracking and recording a person's movements in real time, making it a valuable source of information for emotion recognition. MoCap data from various sub-modes, such as facial, head rotation, and hand, are used in this multimodal task. Despite its relevance, nonverbal signals, including MoCap, have received relatively less attention compared to other modalities.

We collected data points for all feature values within the start and finish time values range and distributed them into 200 separate partitions. These arrays were averaged along the columns to yield 165 significant aspects of the face based on tracking the eyes, eyebrows, nose, mouth, and facial contour. The coordinate information of each point was collected by detecting essential points that express emotional information. The hand modality typically comprises a 2D index with 18 features and a 3D data set with 6 critical points for head rotation. Each sub-mode (facial, hand, and head rotation) was trained separately, and after combining them, a dimensional array of (200, 189) was incurred for each utterance.

CNN-LSTM with Self-Attention architecture was used for emotion classification, where the incurred dimensional vector of (200, 189) was fed into the network. The CNN-LSTM-Attention network comprises 2D convolutional layers with activation, max-pooling, dense layers, and LSTM layers accompanied by a self-attention layer. The CNN layers extracted distinctive features from the motion time series. The convolution operation calculates the convoluted output \textit{z(i, j)} from the input grid value \textit{x(i, j)} and convolution kernel \textit{y(i, j)}. The convoluted output equation is given as:
\begin{equation}
    z\left ( i,j \right) = \sum_{m}^{a-1}\sum_{n}^{b-1}x\left ( m,n\right )\cdot y\left ( i-m, j-n \right )
\end{equation}


where \textit{{a}} and \textit{{b}} are the dimensions of the kernel filter and \textit{z(i, j)} is an element of the two-dimensional convoluted output matrix \textit{{Z}}. The structural architecture of our motion capture is further detailed here. The first convolution layer accepts a two-dimensional array (200, 189) as input. This layer comprises filters with kernel sizes and strides that use the same padding strategy for each convolution operation. The input matrix yields an output matrix of the same size. After the convolution procedure, there is an activation layer, followed by a max-pooling layer then a dropout layer with a rate of 0.2. The dropout layer output is then fed into subsequent convolution blocks of the same structure without the input shape. The procedure is repeated a few more times until the output from the last convolution layer is reshaped and fed into an LSTM block coupled with a self-attention layer.

\subsection{Multimodal Fusion}\label{subsec5}
The fusion process is critical in multimodal speech emotion recognition. A unique fusion process was considered to increase the model's performance in this study. The combination of three modalities (Speech → Sp, Text → Tx, and MoCap → MC) was used to transform the three unimodal systems into a single multimodal emotion system. Four alternative combinations were tested [Sp, Tx], [Sp, MC], [Tx, MC], and [Sp, Tx, MC], and the characteristics of each unimodal phase were extracted independently. The association rule was used to apply the fusion approach. The most powerful speech, text, and MoCap characteristics containing temporal and spatial information were concatenated. Each description can be represented in the following way:\
speech vector $Sp = {\left \{sp_{1}, sp_{2},..., sp_{t}\right \}}$, textual vector $Tx = {\left \{tx_{1}, tx_{2},..., tx_{t}\right \}}$, and MoCap vector $MC = {\left \{mc_{1}, mc_{2},..., mc_{t}\right \}}$, and the final multimodal emotion feature (MEF) vector description being:

\begin{equation}
    MEF = [Sp, Tx, MC]
\end{equation}
In the feature-level fusion or Shallow-Fusion \cite{20} technique used in the multimodal speech emotion recognition paradigm, the concatenated embeddings (MEF) from the three modalities (Speech, Text, and MoCap) were processed by a classification head consisting of a fully connected layer that outputs the relative probability between different emotion classes, followed by a softmax function to capture the relationship between features from different modalities. This fusion technique was chosen for its flexibility and the direct connection between emotional features from multiple modalities and the final decision, as well as for significantly retaining the necessary feature information for the final decision.


\section{EXPERIMENTS}\label{sec4}
\subsection{Dataset}\label{subsec6}
The entire set of experiments was carried out using the publicly available Interactive Emotional Dyadic Motion Capture (IEMOCAP) multimodal and multispeaker database \cite{IE5}, which contains five recorded sessions of conversations (Session 1-5). IEMOCAP is a 12-hour audiovisual dataset with video, speech, motion capture of the face, and text transcriptions rendered by ten professional actors.
To follow existing literature, we chose utterances from only four emotional classes from the database, totaling 5531 utterances. 1708 neutral, 1636 excited (including happy), 1103 angry, and 1084 sad.

\subsection{Setup and evaluation procedure }\label{subsec7}
The experiments used k-fold leave-one-speaker-out cross-validation, with the first four sessions containing 80\% of the data for training and the fifth session containing 20\% of the data for testing. This method ensured speaker independence in the prediction, and a total of 5531 conversations were used in the experiment.

The simulations were done using TensorFlow, and two optimizers, Adam \& Stochastic Gradient Descent, were used for training with a learning rate of 1e-5 and default exponential decay rate of moment estimates. The batch size varied between 4 and 100 during training. The models were trained in a distributed mode using Tesla P100 52GB GPUs. Further details on this will be provided in the upcoming sections.

Our findings are presented in the form of classification accuracy (Equation \ref{eq1}). Furthermore, confusion matrices aided in identifying the errors caused by our classification model during prediction, allowing us to assess the model's validity. The following are provided using the words true positive (TP), false positive (FP), true negative (TN), and false-negative (FN).

\begin{equation}\label{eq1}
    accuracy=  (TP+TN)/(TP+TN+FP+FN)
\end{equation}

\subsection{Training}\label{subsec8}
The experiment's training phase was supervised, and the pre-trained models were trained unsupervised. The model architecture has three modalities, and each was trained separately before combining them for the multimodal task.

\subsubsection{Speech Emotion Recognition Model}\label{subsubsec1}
For speech modality, we experimented by fine-tuning the pre-trained wav2vec2.0-base model with 12 transformer blocks and 7 convolutional blocks, and each block has 512 channels. The model was the backbone of the speech phase and was based on the TensorFlow Hub repository\cite{21}. The fine-tuning ensured that the input was sampled at 16kHz, similar to the pre-trained model. Unlike previous work, which added a Language Modeling (LM) head, the model predicted emotional labels rather than transcribing speech into text. The model signature takes static sequence lengths of 246000, so constants and hyper-parameters were created with an audio max length of 246000. The pre-trained layer was wrapped in successive NN dense layers with ReLU activation, followed by a four-unit classification layer with Softmax activation. The stochastic gradient descent optimizer was utilized with a learning rate of 1e-5, and 1 GPU (Tesla P100) was used for each run with a batch size of 4.

\subsubsection{Text Emotion Recognition Model}\label{subsubsec2}
Fine-tuning BERT for our text phase was simple since the Transformer's self-attention mechanism allowed BERT to simulate downstream tasks by switching out the relevant inputs and outputs. As the backbone for our text, we utilized the BERT base uncased, which contains 12 layers, 12 heads, and 110M parameters. Using the pre-trained BERT, a contextualized word embedding was constructed. Our present model was then given the embeddings. The data is shaped as a 768-dimensional vector with a maximum sequence length of 124, followed by dense layers, and the last layer is a 4-unit output layer with softmax activation. Our model's learning rate was defined as 1e-5 and tuned with Adam optimizer.

\subsubsection{Motion Capture Emotion Recognition Model}\label{subsubsec3}
For the MoCap modality, each sub-modality (facial, head, and hand) was trained separately with up to three deep learning models before being combined into a single motion time series. The head sub-mode included three architectures: model\_1 (2D\_Conv), model\_2 (2D\_Conv-LSTM), and model\_3 (2D\_Conv-LSTM-Attention). All three comprised a stack of 2D convolutional layers, with subsequent max-pooling and dropout layers, followed by dense layers and a final output layer with softmax activation. Adam optimizer with a learning rate of 1e-5 was used in initialization. Model\_2 and Model\_3 included LSTM and LSTM-Attention blocks, which required reshaping the output from the last convolutional layer before feeding into the LSTM blocks. The output from the LSTM and LSTM-Attention blocks was flattened and passed through dense layers. These three models were used for feature learning with the Hand and Facial sub-modes before merging into a single motion time series.

The concatenated motion time series was trained using a 189-feature-length 2D\_Conv-LSTM-Attention architecture. The architecture had convolutional blocks with kernel sizes of 3 and comprised five layers with 128 filters, followed by ReLU activation, max-pooling, and a dropout layer. The output was reshaped and fed into a 128-unit LSTM block coupled with a self-attention layer. The output from the LSTM block was flattened and fed into two dense layers before eventually feeding into a four-neuron dense layer with softmax activation for classification.

\subsubsection{Multimodal Emotion Recognition Model}\label{subsubsec4}
The multimodal phase combined the three modalities assumed in this work. To get the predictions, we concatenated all three vectors at the feature level and sent the concatenated features through a dense layer. Finally, we sent the dense layer output through a classification head, which featured a fully connected layer followed by a softmax function. To comprehend our work better, we investigated additional comparable concatenations such as (Text + Speech, Text + MoCap, and Speech + MoCap). The loss function was the sparse categorical cross-entropy, and Adam was the optimizer. In every iteration, the model was changed in response to the output of the loss function.
Finally, we sent the concatenated embedding through a classification head with a fully connected layer that outputs logits followed by a softmax function.


\section{RESULTS AND ANALYSIS}\label{sec5}
The results of our experiments are summarized in this section. To better understand how important each modality (speech, text, motion capture) was to our multimodal emotion experiment, we trained each modality classifier individually, pairwise, and finally merged all three modalities. Experiments were performed on the IEMOCAP dataset using the leave-one-speaker-out cross-validation, which trained the model during Sessions 1-4 and tested it in Session 5. Our system used emotional audio, text, and MoCap data from IEMOCAP, two task-specific transformer-based models, and a convolutional neural network. The tables below represent the findings and methods utilized in this study.

\begin{table}[h!]
\begin{center}
\caption{Performance comparison of unimodal emotion recognition models on IEMOCAP dataset. The metric used was mean Accuracy(\%) and cross-validation scheme: leave-one-speaker-out (LOSO).}
\begin{tabular}{lccl}
\hline
\textbf{Model}                                                                                                          & \textbf{Validation}                                                                         & \textbf{Accuracy}                                                                                           \\ \hline
\multicolumn{3}{c}{Speech-only}                                                                                                                                                                                                                                                                          \\ \hline
\begin{tabular}[c]{@{}l@{}}LSTM + Attention\cite{B1}\\ GRU\cite{B2}\\ Bi-LSTM\cite{B3}\\ GRU + Attention\cite{11}\\ Bi-LSTM + Attention\cite{B4}\\ \textbf{Sp-Wav2Vec}\end{tabular} & \begin{tabular}[c]{@{}c@{}}Holdout\\ LOSO\\ LOSO\\ LOSO\\ LOSO\\ LOSO\end{tabular} & \begin{tabular}[c]{@{}c@{}}55.65\% \\ 55.92\%\\ 55.60\%\\ 56.65\%\\ 60.83\%\\ \textbf{63.99\%}\end{tabular} \\ \hline
\multicolumn{3}{c}{Text-only}                                                                                                                                                                                                                                                                            \\ \hline
\begin{tabular}[c]{@{}l@{}}LSTM\cite{B1}\\ Bi-LSTM\cite{B3}\\ BERT-GRU\cite{B2}\\ \textbf{Tx-BERT}\end{tabular}                        & \begin{tabular}[c]{@{}c@{}}LOSO\\ LOSO\\ LOSO\\ LOSO\end{tabular}                  & \begin{tabular}[c]{@{}c@{}}64.78\%  \\ 65.90\%\\ 67.24\%\\ \textbf{70.13\%}\end{tabular}                    \\ \hline
\multicolumn{3}{c}{MoCap-only}                                                                                                                                                                                                                                                                           \\ \hline
\begin{tabular}[c]{@{}l@{}}CNN\cite{B1}\\ \textbf{MC-Conv-LSTM-Attention}\end{tabular}                                                             & \begin{tabular}[c]{@{}c@{}}LOSO\\ LOSO\end{tabular}                                & \begin{tabular}[c]{@{}c@{}}51.11\%  \\  \textbf{53.31\%}\end{tabular}                                       \\ \hline
\end{tabular}
\label{table:tab1}
\end{center}
\end{table}

\subsection{Comparison between unimodal inputs in fine-tune state}\label{subsec9}
Table~\ref{table:tab1} illustrates two essential points. It compares the performance of unimodal inputs and fine-tuned approaches on the IEMOCAP dataset. In our experiments, we employed the high-level features of wav2vec and BERT as the sequence representation for our speech and text respectively. As indicated in Table~\ref{table:tab1}, text-only outperformed speech and MoCap exclusively. A possible reason for this result could be the presence of strong emotional cues in the linguistic framework because there are not many intricate circumstances in this dataset where speech is more effective. However, when comparing text-only results to our best-performing multimodal model, we detected a considerable improvement, emphasizing the relevance of multimodality. 

\begin{figure}[h!]
    \centering
    \includegraphics[width=8cm]{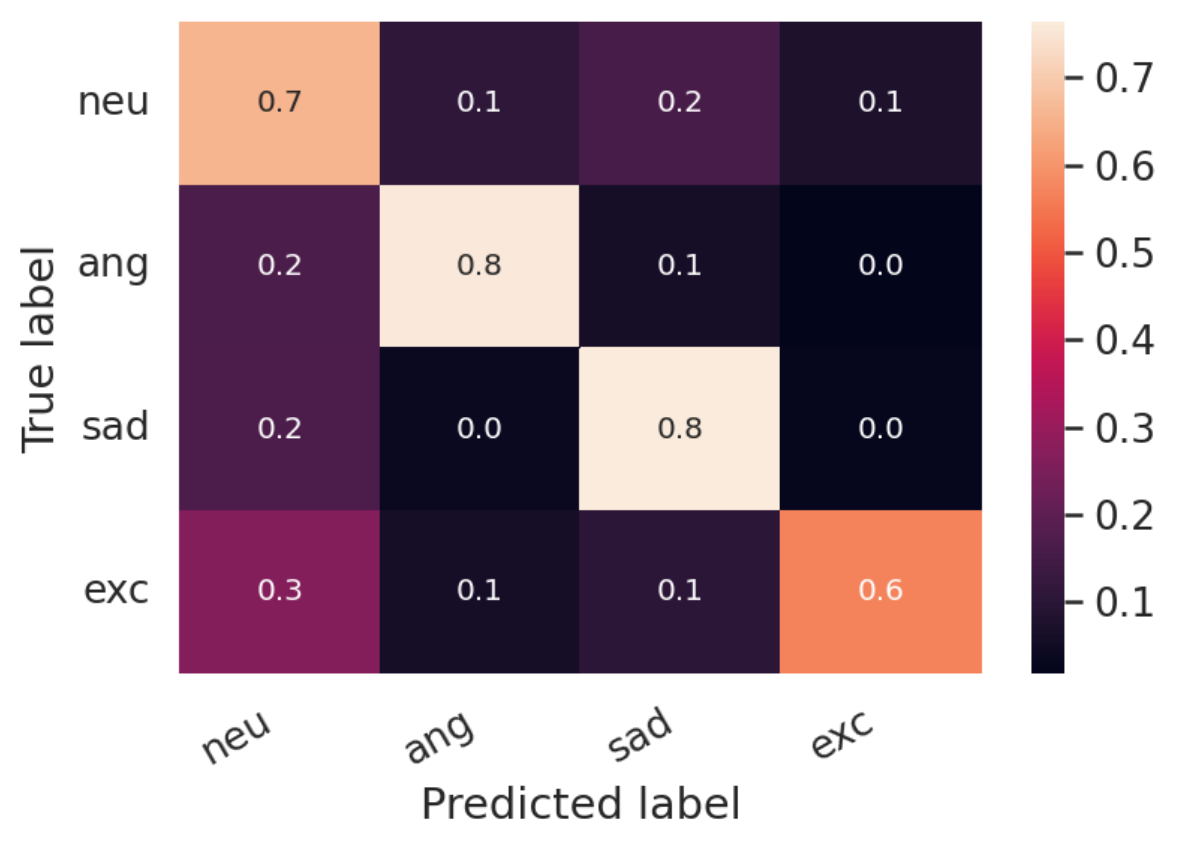}
    \caption{Text\_only normalized confusion matrix on model test accuracy}\label{fig:textonly}
\end{figure}

\begin{figure}[h!]
\begin{center}
    \centering
    \includegraphics[width=8cm]{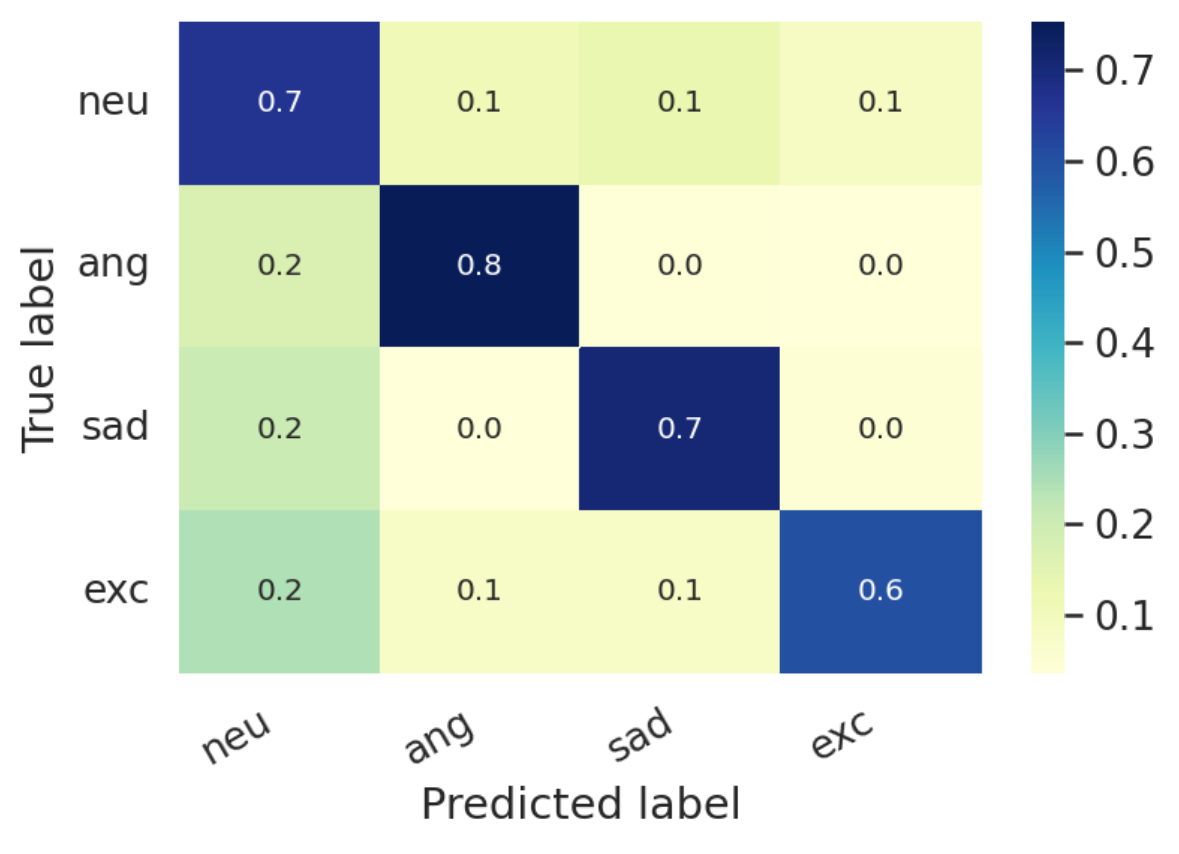}
    \caption{Speech\_only normalized confusion matrix on model test accuracy}
    \label{fig:speechonly}
\end{center}
\end{figure}

We produced the confusion matrices shown in Fig.~\ref{fig:textonly} and~\ref{fig:speechonly} for our unimodal text and speech. In single mode, there was an apparent emotional disorientation. In both voice and text mode, the class {\em excited} was misread as {\em neutral}, with a more significant misjudged prediction percentage of 30 in text mode compared to 20\% in speech. We also observed 70-80\% accuracy prediction in three emotion categories ({\em neutral}, {\em anger}, and {\em sad}), with the {\em excited} category performing the worst in both the speech and text models. The high prediction accuracy observed in the emotional classes might be due to the adaptability of pre-trained architectures employed for feature extraction and prediction in text and speech.  We also observed complementarity between speech and text. As seen in Fig.~\ref{fig:confmx}, their combination improved the accuracy of most emotion categories while reducing misjudgments between emotions.

We also explored whether using wav2vec embeddings for speech emotion recognition had any advantages. We used hyperparameter tweaks to train our speech, such as an Adam optimizer with learning rates of 5e-5 \& 1e-5 and an SGD-optimized architecture with a learning rate of 1e-5. However, we report the findings for the SGD setup, which produced an absolute accuracy of 63.99\% when wav2vec features were used as input. As a result, compared to standard low-level speech features such as Mel-frequency spectrograms, the learning accuracy of the wav2vec embeddings was a superior representation of speech for SER on the IEMOCAP dataset. Compared to the SGD, the Adam-optimized wav2vec model with both learning rates performed marginally.

\begin{table}[]
\begin{center}
\caption{Motion Capture models and accuracy results}
\resizebox{\columnwidth}{!}{%
\begin{tabular}{lcllc}
\hline
\textbf{Method} & \multicolumn{4}{c}{\textbf{Accuracy (\%)}} \\ \hline
 & Hand & \multicolumn{1}{c}{Head} & \multicolumn{1}{c}{Face} & \multicolumn{1}{l}{Combined} \\ \cline{2-5} 
\begin{tabular}[c]{@{}l@{}}Model 1 \\ (CNN)\end{tabular} & 35.43\% & 42.26\% & 49.54\% & -- \\
\begin{tabular}[c]{@{}l@{}}Model 2 \\ (CNN-LSTM)\end{tabular} & \multicolumn{1}{l}{33.70\%} & \textbf{43.35\%} & 44.26\% & -- \\
\begin{tabular}[c]{@{}l@{}}Model 3 \\ (CNN-LSTM-Att)\end{tabular} & \multicolumn{1}{l}{\textbf{36.34\%}} & 42.17\% & \textbf{49.64\%} & \textbf{53.31\%} \\ \hline
\label{table:tab2}
\end{tabular}%
}
\end{center}
\end{table}

The MoCap-only section described concatenating three subsections into a single modality to generate our MoCap input. We trained and evaluated each subsection using several deep learning architectures, and their results are displayed in Table~\ref{table:tab2}. It helped us comprehend the nature of our motion capture data. Since the separate subsections (facial, head, and hand) do not have enough data to operate as independent data points, the results obtained from training each subsection were relatively poor. The combined output exceeded the 2D\_Conv baseline models used in training the separate subsections, demonstrating the efficacy of the proposed 2D\_Conv-LSTM-Attention for the concatenated motion time series over the individual sub-modes of our motion capture data.
\begin{figure}[h!]
\begin{center}
    \centering
    \includegraphics[width=8cm]{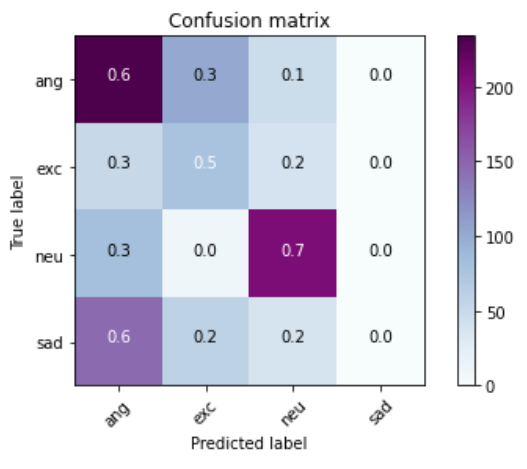}
    \caption{MoCap\_only normalized confusion matrix on model test accuracy}
    \label{fig:mocap_only}
\end{center}
\end{figure}

In our MoCap experiment's confusion matrix, Fig.~\ref{fig:mocap_only}, we discovered that the neutral emotional class had the best-predicted accuracy of 70\%. Despite having the best forecast accuracy, it also recorded a high misjudgment of 30\%. The model failed to predict the {\em sad} emotion among the emotional classes. It also had oscillating predictions throughout all emotional categories, with the anger class coming in second with 60\% prediction accuracy. There was also a 60\% perplexing misrepresentation of {\em sadness} in the anger category. The majority of the downsides experienced are linked to the dataset. We discovered that the Motion Capture markers could not pick up information at several data collecting points, mainly attributed to occlusions or ambiguity. When a significant fraction of markers are missing for lengthy periods, most approaches for filling missing markers may soon become inefficient and yield unacceptable results. The excited class had the lowest prediction accuracy of 50\%.

\begin{table}[ht!]
\begin{center}
\caption{Performance comparison of pairwise approaches and conventional studies on IEMOCAP dataset for four emotional classes.}
\begin{tabular}{lcc}
\multicolumn{1}{l|}{\textbf{}}                                                                                                                                                                & \multicolumn{1}{c|}{\textbf{Modality}}                                                                                                                     & \textbf{Accuracy}                                                                                   \\ \hline
\multicolumn{3}{l}{\textbf{Baseline Methods}}                                                                                                                                                                                                                                                                                                                                                                                                                    \\ \hline
\multicolumn{1}{l|}{\begin{tabular}[c]{@{}l@{}}Liu et al.\cite{B6}\\ Chen et al.\cite{B7}\\ Krishna et al.\cite{B3}\\ Makiuchi et al.\cite{25}\\ Ngoc-Huynh et al.\cite{B2}\\ \textbf{W2V-BERT}\end{tabular}} & \multicolumn{1}{c|}{\begin{tabular}[c]{@{}c@{}}Speech + Text\\ Speech + Text\\ Speech + Text\\ Speech + Text\\ Speech + Text\\ Speech + Text\end{tabular}} & \begin{tabular}[c]{@{}c@{}}70.08\% \\ 72.05\%\\ 72.82\%\\ 73.00\% \\ 73.23\%\\ \textbf{74.93\%}\end{tabular} \\ \hline
\multicolumn{3}{l}{{Proposed Pairwise Versions}}                                                                                                                                                                                                                                                                                                                                                                                                             \\ \hline
\multicolumn{1}{c|}{\textbf{Bimodal Fusions}}                                                                                                                                                            & \multicolumn{1}{c|}{\begin{tabular}[c]{@{}c@{}}Speech + Mocap\\ Text + MoCap\end{tabular}}                                                                 & \begin{tabular}[c]{@{}c@{}}68.02\%\\ 71.77\%\end{tabular}                                           \\ \hline
\end{tabular}
\label{table:tab3}
\end{center}
\end{table}

\subsection{Comparison of the pairwise fusions}
Table~\ref{table:tab3} displays the results of our pairwise fusion model with various modality combinations. In contrast to the unimodal results presented in Table~\ref{table:tab1}, combining any pairings can substantially enhance performance. It shows a considerable complementary nature of the two modes on this task. Table~\ref{table:tab3} further shows that the Speech + Text feature-level fusion approach outperforms the other fused models (speech + MoCap and Text + MoCap). When combining the speech and text findings, placing less emphasis on the speech model is acceptable because the text model compensates for the majority of emotional perplexity in the combined phase. We hypothesize that this is due to the certainty with which the text model received its scores. From Table~\ref{table:tab3}, we can also see that all paired combinations containing the text modality obtained a result significantly higher than the pairwise version (speech + MoCap) that did not include the text. It is possible to conclude that merging distinct modalities contributes sensibly to the emotion recognition task because our pairwise findings outperformed our speech-only, text-only, and MoCap-only results.
\begin{table}[h!]
\begin{center}
\caption{Performance comparison between our proposed model and state-of-the-art(SOTA) multimodal model on the IEMOCAP dataset, using three modalities.}
\begin{tabular}{lccl}
\hline
                        & \textbf{Method}                           & \textbf{Modality}                         & \textbf{Accuracy} \\ \hline
Tripathi et al.\cite{B1}      & LSTM+CNN+Att                    & Sp+Tx+MC                             & 71.04\%           \\
\textbf{Proposed Model} & \multicolumn{1}{l}{\textbf{W2V-BERT-CLA}} & \multicolumn{1}{l}{\textbf{Sp+Tx+MC}} & \textbf{77.58\%}  \\ \hline
\end{tabular}
\label{table:tab4}
\end{center}
\end{table}


\subsection{Comparison between State-of-the-Art (SOTA) methods and the Proposed Approach}\label{subsec10}
\begin{figure}[h!]
\vspace{3mm}
\begin{center}
    \centering
    \includegraphics[width=8cm]{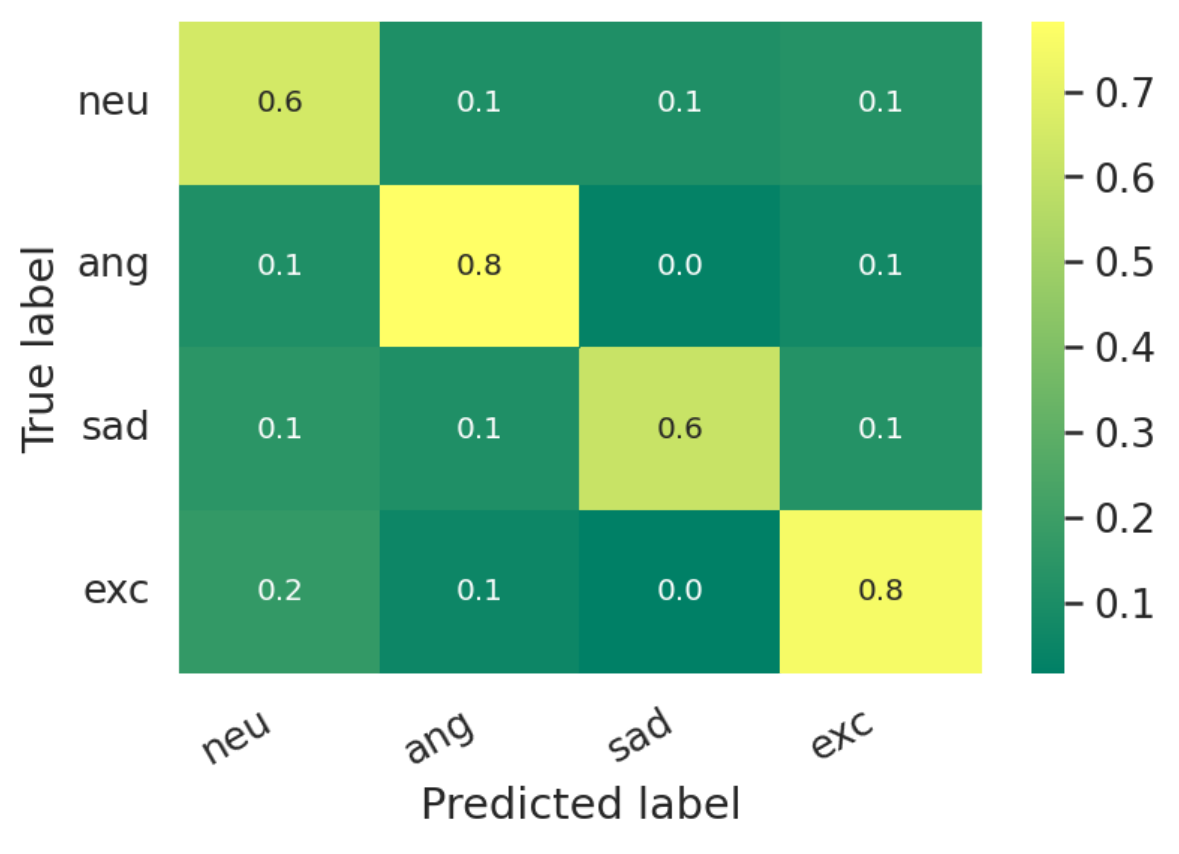}
    \caption{Multimodal normalized confusion matrix on model test accuracy}
    \label{fig:confmx}
\end{center}
\end{figure}

This section further evaluates the efficacy of the proposed approach, showing that combining the uni-modal models at the feature level can improve performance. Table~\ref{table:tab4} compares the proposed approach with other state-of-the-art multimodal (speech-text-mocap) emotion recognition approaches. To ensure equitable comparison, all the approaches are based on the leave-one-speaker-out (LOSO) cross-validation configuration, using the mean accuracy as the metric and analogous preprocessing of the IEMOCAP data as executed in this work. These results show that the performance of our model, using speech, text, and MoCap features, was superior to that of most baseline approaches, suggesting that it is advantageous to combine wav2vec 2.0, BERT, and CNN-LSTM with a Self-Attention module. It is worth noting that the improvement of our mode is not limited to the effect of the fusion approach. Furthermore, an added contributor that influenced the result is the integration of prosodic data, which holds significance in the recognition of emotions. In addition, our multimodal model was evaluated using a produced confusion matrix shown in Fig.~\ref{fig:confmx}.

The multimodal model has a more significant recognition accuracy for {\em anger} and {\em excited} and lower recognition accuracy for {\em neutral} and {\em sad}. Because the accuracy results obtained on our unimodal models in Table~\ref{table:tab1} above are marginal, we argue that our multimodal model compensates for and benefits from the deficiencies of the single modalities. Despite extensive emotion recognition capabilities, the multimodal model made a few mistakes, misclassifying 20\% of the emotional excitement class as neutral and 30\% as a split between anger, sadness, and excitement. This might be due to some disparity in our dataset since some specific emotional categories had more data than others, or it could have been due to the simple fusion type. We intend to investigate different fusion approaches in the future to see how they differ.

\section{CONCLUSION}\label{sec6}
This paper presented a multimodal emotion recognition system that addresses the downstream issue of multimodal speech recognition using modality-specific pre-trained structures and nonverbal behavioral cue data points acquired through motion capture. We fine-tuned Wav2vec 2.0 and BERT architectures for speech and text modalities, respectively, and utilized three sub-modes for the MoCap modality. The resulting system significantly improves emotion recognition by fusing unimodal modality information.

We experimented with individual and paired modalities before concatenating the three modalities (speech, text, and MoCap) at the feature level, which improved the performance of their emotion recognition system and reduced data sparsity. Merging the modalities compensated for one another's flaws and enhanced the system's performance while reducing data sparsity. The best fusion configuration achieved a mean accuracy of 77.58\% on the IEMOCAP dataset for leave-one-speaker-out (LOSO) cross-validation.

In the future, we intend to explore a pre-trained approach for the motion capture data point to avoid training from scratch, which requires excessive computational resources. We aim to use the pre-trained model's advantage to extract more valuable emotion features from MoCap, combined with the two architectures for speech and text used in the study, to improve the multimodal model's performance. Other fusion strategies that best suit features extracted from different pre-trained architectures will also be considered. Additionally, cross-database validation will be used to assess the methods' resilience further.

\bibliographystyle{IEEEtran}
\bibliography{citation.bib}

\end{document}